\documentclass[twocolumn,groupedaddress,amsmath,amssymb,prb,aps,longbibliography]{revtex4-1}

\usepackage{graphicx}% Include figure files
\usepackage{dcolumn}% Align table columns on decimal point
\usepackage{bm}% bold math
\usepackage{booktabs}

\usepackage{tikz}
\definecolor{mydarkgreen}{rgb}{0.0, 0.5, 0.0} % RGB for dark green
\usepackage{rotating}

\newcommand{\bto}{BaTiO$_3$}
\newcommand{\bfo}{BiFeO$_3$}
\newcommand{\pzo}{PbZrO$_3$}
\newcommand{\hfo}{HfO$_2$}

\begin{document}

\title{Minimalist machine-learned interatomic potentials can predict
  complex structural behaviors accurately}

\author{I\~nigo Robredo-Magro,$^{1}$ Binayak Mukherjee,$^{1}$ Hugo
  Aramberri,$^{1}$ and Jorge \'I\~niguez-Gonz\'alez$^{1,2}$}

\affiliation{
  \mbox{$^{1}$Smart Materials Unit, Luxembourg Institute of Science
    and Technology (LIST),}
  {Avenue des Hauts-Fourneaux 5, L-4362 Esch/Alzette, Luxembourg}\\
 \mbox{$^{2}$Department of Physics and Materials Science, University
   of Luxembourg, Rue du Brill 41, L-4422 Belvaux, Luxembourg}}

\begin{abstract}
The past decade has witnessed a spectacular development of
machine-learned interatomic potentials (MLIPs), to the extent that
they are already the approach of choice for most atomistic simulation
studies not requiring an explicit treatment of electrons. Typical MLIP
usage guidelines emphasize the need for exhaustive training sets and
warn against applying the models to situations not considered in the
corresponding training space. This restricts the scope of MLIPs to
interpolative calculations, essentially denying the possibility of
using them to discover new phenomena in a serendipitous way. While
there are reasons to be cautious, here we adopt a more sanguine view
and challenge the predictive power of two representative and widely
available MLIP approaches. We work with minimalist training sets that
rely on little prior knowledge of the investigated materials. We show
that the resulting models -- for which we adopt modest/default choices
of the defining hyperparameters -- are very successful in predicting
non-trivial structural effects (competing polymorphs, energy barriers
for structural transformations, occurrence of non-trivial topologies)
in a way that is qualitatively and quasi-quantitatively correct. Our
results thus suggest an expanded scope of modern MLIP approaches,
evidencing that somewhat trivial -- and easy to compute -- models can
be an effective tool for the discovery of novel and complex physical
phenomena.
\end{abstract}

\maketitle

%\tableofcontents

\section{Introduction}

Machine-learned force fields or interatomic potentials (MLFFs or
MLIPs) have become an efficient tool for first-principles atomistic
modeling.\cite{bartok15,wang18,jinnouchi19,fan21,musaelian23,xie23,batatia24,jacobs25}
When trained on quantum-mechanical data, they can achieve
near-first-principles accuracy at a small fraction of the cost. Most
current applications of MLIPs rely on large, carefully selected
training sets and optimized hyperparameters, with the aim of ensuring
that the model interpolates within its training domain. This strategy
minimizes errors and yields very accurate potentials. It also seems to
imply that the scope of MLIPs is confined to regions of the
configuration space explored in the training set.

Ferroelectric materials exhibit highly non-trivial structural and
lattice-dynamical behavior driven by competing instabilities and strong
anharmonic
effects.\cite{rabe-book2007,meier-book2020,junquera23,schroeder-book2025}
Their phase diagrams often include multiple polymorphs with
quasi-degenerate free energies, intricate transformation paths between
phases, and complex topological features. Further, their properties
are highly sensitive to electric and elastic boundary conditions,
size, temperature, and external fields. First-principles methods, as
those based on Density Functional Theory
(DFT),\cite{hohenberg64,kohn65,martin-book2004} can predict and
explain such effects with high accuracy, but their computational cost
limits their use in large-scale or long-time simulations. Successful
partial solutions to this problem have been developed over the years,
notably through the so-called ``second-principles''
methods\cite{wojdel13,escorihuelasayalero17,garciafernandez16,carralsainz25}
and other first-principles-based effective potential
approaches.\cite{zhong94a,zhong95a,waghmare97,krakauer99,bellaiche00a,prosandeev13,zubko16,brown09,shin05,tinte99}
Yet, such schemes lack generality, which limits the span of materials
and effects investigated. For example: novel fluorite ferroelectrics
such as hafnia, characterized by an intimate connection between
ferroelectric switching and ionic mobility, are all but impossible to
treat within the traditional effective schemes. Modern MLIPs are
expected to drastically improve this situation and become the approach
of choice for simulations of ferroelectric and related
phenomena. Symbiotically, ferroelectrics offer a stringent playground
for testing the performance of modern ML approaches.

In this context, it is relevant to note that relatively simple
second-principles potentials and effective Hamiltonians have been
remarkably successful in {\it predicting} behaviors beyond the scope
of the DFT data used to compute the model parameters. Notable examples
include complex composition-driven phase transitions (as in the
so-called ``morphotropic phase boundary'' of
PbZr$_{1-x}$Ti$_{x}$O$_{3}$ solid solutions\cite{bellaiche00a}) or the
occurrence of topological skyrmion-like quasiparticles in
nanoferroelectrics.\cite{naumov04,nahas15,goncalves19,das19,aramberri24}
These and other striking discoveries -- eventually confirmed
experimentally -- emerged serendipitously from the simulations. Can
MLIPs deliver similarly unforeseen predictions, or are these the
patrimony of relatively simple models that can be expected to yield
meaningful results beyond their training space? This is a relevant
question to define the application scope of MLIPs and to what extend
resorting to simpler potentials may still be required in exploratory
studies.

In this work we test the predictive performance of MLIPs for
challenging ferroelectricity-related effects and provide an answer to
the above question. We adopt a peculiar approach, whereby we {\sl
  pretend} we know little about the materials at the time of
constructing the models; thus, we use minimalist training sets of the
type one would consider by default as a first step in the
investigation of an all-new compound. Then, we test whether the MLIPs
thus constructed are able to predict the non-trivial structural
properties that we know these compounds display.

Specifically, we work with two widely available and representative
MLIP types: the kernel-based Gaussian Approximation
Potentials\cite{bartok15} (GAPs) implemented in the {\sl Vienna ab
  initio software package}\cite{jinnouchi19,kresse96,kresse99} (VASP)
and the equivariant deep learning potentials provided in {\sl
  Allegro}.\cite{musaelian23,tan25,batzner22} In both cases we adopt
modest or default (when available) choices for the hyperparameters
defining the MLIPs, which are supposed to strike a balance between
accuracy and computational burden in the description of generic
compounds. Also, we use the Bayesian ``on-the-fly'' approach
implemented in VASP to derive training sets,\cite{jinnouchi19} using
material-unspecific control parameters and starting the procedure from
the experimentally well-known structural phases of the respective
compounds. Following this minimalist spirit, we generate our training
sets using small periodically-repeated supercells containing 40 atoms
at most.

We apply this MLIP-construction approach to four representative
ferroelectric and related compounds, namely BaTiO$_{3}$, BiFeO$_{3}$,
PbZrO$_{3}$, and HfO$_{2}$. We find that our basic models successfully
reproduce a variety of non-trivial behaviors, such as the
vortex-antivortex electric-dipole lattice recently observed in {\bto}
moir\'e bilayers,\cite{sanchezsantolino24} the non-trivial
polarization switching path (and its corresponding energy barrier) in
{\bfo},\cite{heron14} the competition of polar and antipolar
polymorphs in {\pzo},\cite{iniguez14,aramberri21} and the occurrence
of a ferroelectric polymorph in {\hfo}.\cite{boscke11} In all cases we
find that the models predict the examined properties in a
qualitatively correct way. Furthermore, our quantitative results --
including tiny energy differences between competing polymorphs and
polarization-switching energy barriers -- turn out to be incredibly
accurate in most cases.

The structure of the paper is as follows. In Section~\ref{sec:mlip} we
describe the construction of our minimalist MLIPs. In
Section~\ref{sec:results} we present the obtained models and how well
they perform as regards both interpolating within the training set and
extrapolating to unexplored space. In Section~\ref{sec:discussion} we
further discuss the implications of this study. Finally, in
Section~\ref{sec:summary} we summarize our main findings and
conclusions.

\section{Minimalist models}\label{sec:mlip}

The guiding principle of our modeling approach is simplicity and
minimal prior knowledge. For all models considered in this work, we
deliberately adopt a low-effort protocol: training sets are kept small
are left at their default values; no optimization or manual tuning is
performed at any stage. This strategy allows us to evaluate the
intrinsic predictive capacity of modern MLIP frameworks under the most
basic by-default fool-proof conditions. Our aim is not to benchmark
best-case performance, but rather to assess how well these models can
generalize when trained in a straightforward, out-of-the-box manner.

\subsection{Kernel-based GAP interatomic potentials}

We first consider a kernel-based interatomic potential trained using
the Bayesian on-the-fly framework implemented in
VASP.\cite{bartok15,jinnouchi19} This method is rooted in kernel ridge
regression, using Smooth Overlap of Atomic Positions (SOAP)
descriptors\cite{bartok13} to represent local atomic environments and
build GAPs that mimic DFT accuracy. Training data is gathered during
relatively short first-principles molecular dynamics (MD) simulations
of the studied systems.\cite{jinnouchi19} We train starting from
well-known polymorphs of each of the compounds, running for 10000
steps of 2~fs, with a linear temperature ramp from 10~K to 300~K.

We set the model parameters to the default values of VASP: the cut-off
radius for the radial descriptor is 8~\AA\ (which roughly coincides
with the lattice parameter used for the perovskite supercells), the
cut-off radius for the angular descriptor is 5~\AA\ (enough to include
at least 20 nearest neighbors for each atom). The number of basis
functions used to expand the radial and angular descriptors is set to
12 and 8, respectively, and the maximum angular momentum of spherical
harmonics is set to 6. The threshold for error estimation is set to
0.002, with automatic updating to a value proportional to the average
errors as implemented by default. As the simulation runs, the model
learns in real time: in essence, new configurations are flagged if the
predicted maximum Bayesian error is larger than a threshold, and a new
first principles calculation is performed, adding it to the dataset.
In the following, we refer to these models generically as ``VASP
MLIPs''.

\subsection{E(3)-equivariant neural networks}

We also constructed potentials based on the Allegro neural network
architecture,\cite{musaelian23,tan25,batzner22} trained using the very
same dataset generated by the VASP on-the-fly procedure.  Allegro is
based on the same symmetry principles as the better-known NequIP
scheme -- specifically, E(3)-equivariance -- but differs in how it
processes atomic environments. Instead of relying on message passing
between atoms, Allegro builds local representations directly from the
fixed neighborhood of each atom, using a sequence of tensor products
to capture many-body interactions. This design avoids the iterative
communication steps typical of message-passing networks, leading to
improved scalability and faster evaluation, particularly in large
systems.

We make a reasonable choice of Allegro hyperparameters based on the
documentation provided by the developers, with no effort to tune
them. We set the radial cut-off to the same value as for the VASP
MLIPs, 8~\AA. For the rest of the parameters, we take them as in the
application example proposed with Allegro. We use 2 tensor layers with
full O(3) symmetry, polynomial cut-off of 6, and embed feature
multiplicity of 64. We set the maximum order of spherical harmonics
embedding to 3, which is the recommended value for more accurate MLIPs
while being slower. We train on energies, forces, and stresses, with a
relative error factor of 1:100:100. We split training and validation
sets in a ratio of 9:1 (common ratio) and use a batch size of 1. We
start with a learning rate (LR) of $10^{-3}$ and use a LR scheduler
that reduces it by half on plateaus with a patience of 50 epochs,
stopping the training when the LR is smaller than $10^{-5}$. In the
following, we refer to these models generically as ``Allegro MLIPs''.

\begin{figure}[t]
    \centering
    \includegraphics[width=1\linewidth]{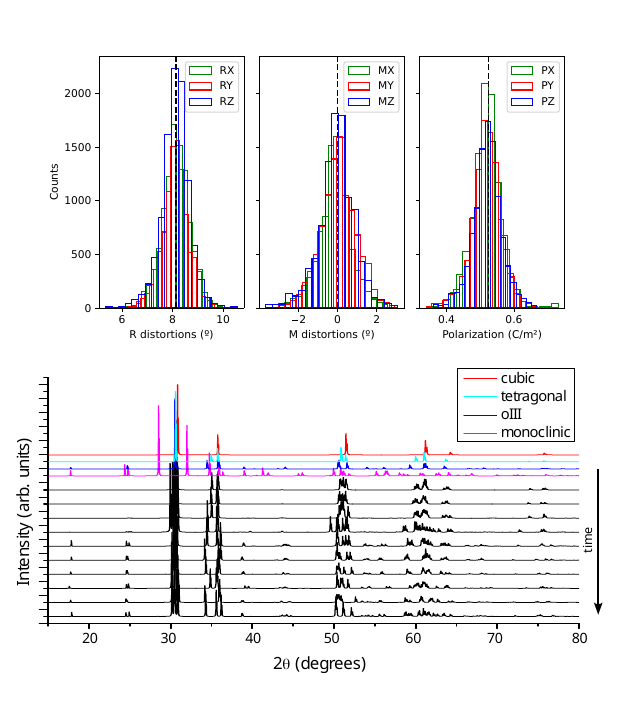}
    \caption{Configuration space explored in the training of {\bfo}
      and {\hfo}. The top panels show histograms for the main
      perovskite distortions: antiphase ($R$) and in-phase ($M$)
      octahedral tilts, as well as the polarization ($P$). The
      histograms are derived from the MD trajectories corresponding to
      the on-the-fly training starting from the ferroelectric ground
      state ($R3c$ polymorphs); they show that the material fluctuates
      around the ground state configuration, characterized by
      $R_{x}=R_{y}=R_{z}\neq 0$, $P_{x}=P_{y}=P_{z}\neq 0$, and
      $M_{x}=M_{y}=M_{z}=0$ (values indicated by dashed black
      lines). The bottom panel shows simulated diffraction patterns
      for configurations visited during the {\hfo} training,
      corresponding to the MD run that started from the tetragonal
      polymorph. By comparison with reference data (colored) for
      important {\hfo} phases, we can appreciate that the material
      evolved spontaneously toward low-symmetry polymorphs during the
      MD run, visiting the oIII ferroelectric phase in particular.  }
    \label{fig:expspac}
\end{figure}

\subsection{On-the-fly training sets} 

In this minimalist approach we limit the number of training
runs. Specifically, for {\bto} we only train once, starting from the
ferroelectric $R3m$ ground state of the material. For both {\bfo} and
{\pzo}, we train first starting from the $R3c$ ferroelectric polymorph
(which is the ground state for {\bfo}), and then we expand the
training set via a second MD run starting from the $Pnma$ non-polar
polymorph.\cite{dieguez11,iniguez14} Finally, for {\hfo} we train
first starting from the $P2_1/c$ non-polar monoclinic ground state, to
then extend the training set with a second MD run starting from
another non-polar polymorph with tetragonal symmetry $P4_{2}/nmc$. For
the perovskites ({\bto}, {\bfo}, and {\pzo}) we always train using a
40-atom supercell that can be viewed as a $2\times 2\times 2$ multiple
of the elemental 5-atom perovskite cell. Note that, in particular,
this implies that the training for {\pzo} never explores the
lowest-energy antiferroelectric and ferrielectric polymorphs of the
material.\cite{aramberri21} For {\hfo}, we train using a 24-atom that
can be seen as obtained from the primitive cell of the tetragonal
polymorph by applying a transformation that doubles the volume while
keeping the cell as isotropic as possible.

In Fig.~\ref{fig:expspac} we illustrate two methods to track the
explored space during the training. For perovskites, since we are
training in the 40-atom unit cell, we can extract the amplitude of the
main distortions that characterize the structures, namely the
polarization ($P$) and the in-phase ($M$) and antiphase ($R$) tilts of
the O$_{6}$ octahedra. In the top panels of Fig.~\ref{fig:expspac} we
show the histogram of occurrences of these distortion along the
training trajectory for {\bfo} starting from the $R3c$ phase. We can
see thermal activation makes the system fluctuate around the ground
state. For {\hfo}, the structural complexity makes it all but
impossible to work with a reduced number of key distortions. Instead,
as shown in the bottom panel of Fig.~\ref{fig:expspac}, we compute the
powder diffraction spectrum (using the visualization software
VESTA\cite{momma11}) of snapshots taken from the training MD
trajectory and compare them to the result for well-known polymorphs,
namely non-polar cubic, tetragonal, and monoclinic as well as the
polar orthorhombic phase (oIII). This MD trajectory corresponds to the
second stage of our {\hfo} training, which starts from the tetragonal
phase with no diffraction peaks present below 30 degrees. However,
midway through the training, low-angle peaks appear at around 18 and
25 degrees. These peaks can correspond to either the oIII or the
monoclinic phase -- however, the characteristic monoclinic peak below
30 degrees is missing, confirming the spontaneous occurrence of the
oIII phase in the MD trajectory.

The training sets thus generated contain the following number of
structures for which a DFT calculation was performed: 247 for {\bto},
488 for {\bfo}, 1300 for {\pzo}, and 463 for {\hfo}. Let us stress
that these are very small numbers, as typical ML models in the
literature rely on training sets at least 10-times
bigger. Representative examples are the deep neural networks recently
proposed for {\pzo} in Ref.~\onlinecite{zhang24} (based on over 12500
structures and a very carefully curated exploration of the energy
landscape, probably including supercells of up to 80 atoms) or for
{\hfo} in Ref.~\onlinecite{wu21} (96-atom supercell, 21500
structures). Naturally, we do not claim here that our minimal models
will be as accurate or complete as the mentioned potentials. Yet, as
we will show, it is worth examining how useful they may be.

\subsection{Details of the first-principles calculations} 

All first-principles calculations were performed using the projector
augmented-wave (PAW) method\cite{blochl94} as implemented in
VASP,\cite{kresse96,kresse99} with the PBEsol exchange-correlation
functional.\cite{perdew96} We solve explicitly for 6 electrons of O
(2s${}^2$p${}^4$), 10 of Ba (5s$^2$p$^6$6s$^2$), 12 of Ti
(3s$^2$p$^6$4s$^2$3d$^2$), 15 of Bi (5d${}^{10}$6s${}^2$6p${}^3$), 14
of Fe (3p${}^6$3d${}^7$4s${}^1$), 14 of Pb (5d$^{10}$6s$^2$p$^4$), 12
of Zr (4s$^2$p$^6$5s$^2$4d$^2$) and 10 of Hf (5p$^6$5d$^3$6s$^1$). The
plane-wave energy cutoff was set to 500~eV in all cases. For Brillouin
zone integrations corresponding to the 40-atom cell in our perovskite
simulations, we use a $\Gamma$-centered $3\times 3\times 3$ $k$-point
mesh for {\bto} and {\pzo}, and a $2\times 2\times 2$ mesh for
{\bfo}. For {\hfo}, we use a grid of $3\times 3\times 3$ $k$-points
for the simulations of the considered 24-atom supercell. For {\bfo},
we add a Hubbard~$U$ correction of 4~eV for a better treatment of
iron's $3d$ electrons.\cite{dudarev98}

\section{Results}\label{sec:results}

We now present our minimalist MLIPs and test their performance. We
start by considering how well the models describe configurations that
appear during the first-principles MD runs used to derive the
corresponding training sets. We thus examine the models' ability to
interpolate between structures typical of the explored space, where
they are expected to perform very well. Then, we run extrapolation
tests designed to push the models outside their comfort zone. Here we
look at structures and properties well beyond the training set,
including features that are critical in investigations of phase
transitions (e.g., dynamic stability of higher-symmetry reference
phases, stability and energy of additional polymorphs), non-linear
responses to external fields (switching paths and energy barriers),
and emerging strongly inhomogeneous orders (nanoscale topological
textures). These tests allow us to gauge to what extent the models can
generalize and still produce physically meaningful results, even when
applied in regimes they were not trained for.

\begin{figure*}[t]
    \centering
    \includegraphics[width=\linewidth]{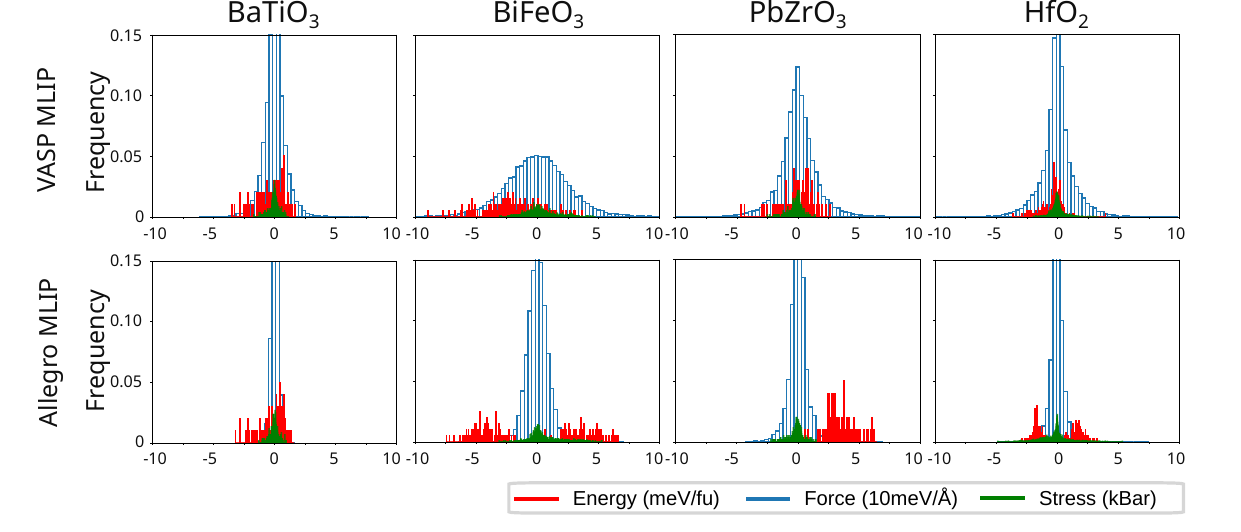}
    \caption{Histograms for the errors made by our minimalist MLIPs in
      reproducing the DFT energies, forces, and stresses corresponding
      to structures typical of the explored configuration space (see
      text). Each distribution is normalized to 1. We use different
      bin sizes for clarity.}
    \label{fig:histograms_main}
\end{figure*}

\subsection{Interpolation tests}

To quantify our models' accuracy at interpolating inside the training
space, we proceed as follows. From each of the MD simulations used to
generate training sets on the fly, we extract 1\% of the visited
structures, equispaced. For each such structure we run a single-point
DFT calculation to compute energy, forces, and stresses. We run the
same calculations using the corresponding MLIPs and compare their
predictions to the DFT results. The histograms of the differences
between MLIP predictions and DFT results are shown in
Fig.~\ref{fig:histograms_main}.

We find that, across all systems, both our VASP and Allegro models
show relatively narrow error distributions, which implies that the
MLIPs perform well reproducing DFT results. Most of the predictions
fall within $\pm$5~meV per formula unit (f.u.) for energy,
$\pm$50~meV/{\AA} for force components, and $\pm$3~kBar for stress
components. Both models show a broader and noisier error distribution
for the energies, suggesting that they are harder to learn. This could
be due to the relatively small number of energy data points (1 per
configuration in the training set) as compared to forces (tens per
configuration) or stresses (6 per configuration). Among the considered
materials, the models for {\hfo} seem the most accurate. Overall,
while both approaches work well to reproduce our minimal datasets,
Allegro seems to offer a small edge in accuracy within the explored
configuration space.

\subsection{Extrapolation tests}

We perform two kinds of extrapolation tests. First, for all the
materials we compute the phonons of a high-symmetry phase that appears
as the natural reference from which all the low-energy structures of
interest can be obtained through appropriate distortions. This
reference phase is the ideal cubic perovskite structure for {\bto},
{\bfo}, and {\pzo}, and the ideal cubic fluorite structure for {\hfo}.
Note that the phonons of such reference structures provide a guide to
the symmetry- and energy-lowering distortions of the materials, and
are thus critical in the study of phase transitions. (The physical
relevance of the cubic fluorite structure to ferroelectric {\hfo} is
currently debated,\cite{aramberri23} but considering the phonons of
this phase still serves our current testing purposes.)

Second, we perform a battery of material-specific tests. For {\bto} we
examine the vortex-antivortex electric dipole texture recently
reported in Ref.~\onlinecite{sanchezsantolino24}, one of the
foundational results of the field of moir\'e multilayers based on
perovskite-oxide membranes. For {\bfo} we examine the non-trivial
structural transition path associated to the reversal of one
polarization component, key to the magnetoelectric switching
properties (i.e., electric control of the magnetization) in this
material.\cite{heron14,fedorova24} For {\pzo} we inspect the relative
stability of low-lying polymorphs of diverse electric character
(antiferro, ferro, ferri), critical to the phase transitions and
antiferroelectric response of this compound in bulk and thin-film
forms.\cite{aramberri21,parmar25} Finally, for {\hfo} we examine the
existence and relative stability of the ferroelectric polymorph that
has brought massive interest to this compound, including some of its
well-known distinctive features (e.g., the existence of domain walls
with a negative formation
energy).\cite{boscke11,lee20,azevedoantunes22,aramberri23}

\begin{figure*}[t]
    \centering
    \includegraphics[width=\linewidth]{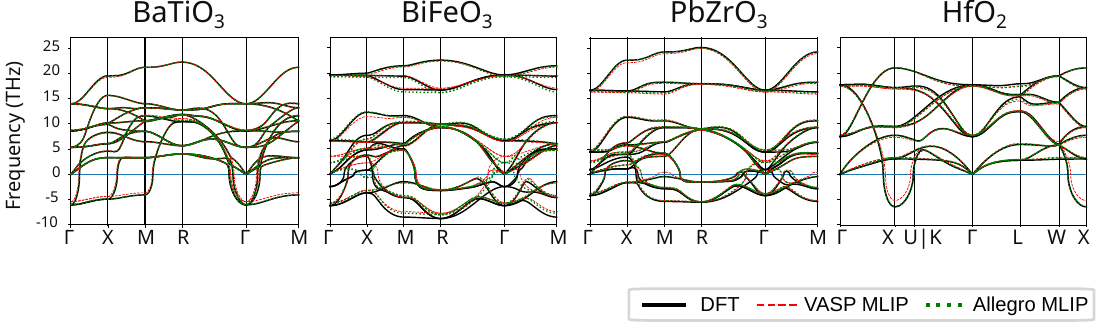}
    \caption{Phonon bands for high-symmetry reference phases (see
      text). In black, solid lines, the DFT result. In red, long
      dashed lines, the VASP MLIP prediction. In green, short dashed
      lines the Allegro MLIP prediction.}
    \label{fig:phonons}
\end{figure*}

\subsubsection{Reference-phase phonon bands}

We compute the phonons of the cubic phase of each material using DFT
and the corresponding MLIP, and compare the results in
Fig.~\ref{fig:phonons}. (We employ the package phonopy \cite{phonopy}
to compute phonons, using the finite-differences approach in a
$2\times 2\times 2$ supercell. For the sake of a more direct
comparison, we omit non-analytical corrections in these calculations.)
Overall, our minimalist MLIPs capture the main features of the phonon
spectra remarkably well, reproducing exactly the DFT results for most
bands. Notably, all MLIPs correctly predict the main structural
instabilities. In {\bto}, for example, the models capture the unstable
bands with imaginary phonon frequencies (shown as negative frequencies
in Fig.~\ref{fig:phonons}, following the usual convention). The most
notable disagreement is found for {\bfo}, where the MLIPs do not
capture a second unstable phonon at the $\Gamma$ point; this is a fine
detail that does not have -- as far as we can tell -- any practical
consequence.

Remarkably too, our MLIPs do not seem to introduce any unphysical
features. Most critically, we find no new unstable modes that could
lead to artifacts such as false energy minima (i.e., fake competing
polymorphs) or qualitative errors (fake anomalies) in the response to
external fields. This strongly suggests that our minimal MLIPs are
both robust and reliable, predicting behaviors that are physically
sound even outside the explored training space.

\begin{figure*}[t]
    \centering
    \includegraphics[width=1\linewidth]{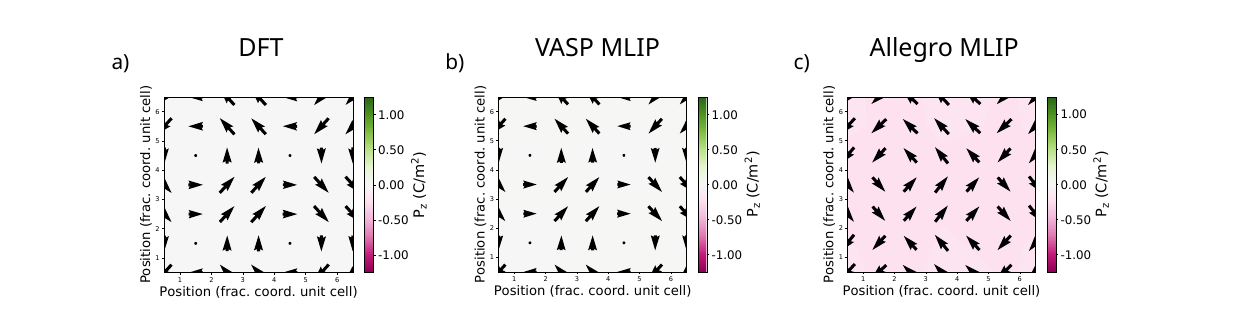}
    \caption{Vortex-antivortex electric dipole patterns mimicking
      experimental observations in {\bto} moir\'e bilayers (see
      text). The arrows denote in-plane local polarization components
      (see text), while the color map corresponds to the out-of-plane
      component.}
    \label{fig:bto}
\end{figure*}

\subsubsection{BaTiO$_{3}$: Vortex-antivortex electric texture}

Bulk barium titanate is paraelectric and cubic above 393~K,
approximately.\cite{merz49,merz49b,lines-book1977} Upon cooling it
undergoes a well-known sequence of phase transitions to ferroelectric
phases of different symmetry: tetragonal, orthorhombic, and
rhombohedral. In ideal bulk-like conditions (i.e., electric short
circuit and zero remnant stress), the polar states tend to be
homogeneous. Under specific electric or elastic boundary conditions --
e.g., as in epitaxial thin films -- {\bto} presents multidomain states
that essentially follow well-known laws such as
Kittel's\cite{kittel46}.

Recently, experiments on twisted freestanding bilayers of {\bto} have
revealed a mesmerizing and unprecedented type of ferroelectric
multidomain configuration: a lattice of electric vortices and
antivortices that appears at the interface between
layers~\cite{sanchezsantolino24}, essentially as shown in
Fig.~\ref{fig:bto}. Costly DFT calculations helped ascertain the
origin of such dipole texture: the shear-strain modulation induced by
the moir\'e interfacial potential gives rise to the observed vortical
pattern through the flexoelectric effect, i.e., the polar response to
strain gradients.\cite{zubko13} The DFT analysis concluded that the
vortex-antivortex dipole lattice lies about 9~meV/f.u. above an
homogeneous state with a net polarization along a $\langle 110
\rangle$ in-plane pseudocubic direction. The simulations relied on
drastic simplifications -- e.g., a bulk-like version of the
vortex-antivortex lattice was considered, instead of the actual
moir\'e bilayer. Nevertheless, the connection between polar texture
and inhomogeneous strain was clearly established, evidencing a key
role of the flexoelectric effect that was corroborated by other
authors shortly after~\cite{shahed25,prosandeev25}.

The mentioned vortex-antivortex lattice displays many non-trivial
features that provide a stringent test to the predictive power of our
minimalist MLIPs for {\bto}. Most notably, it displays strong
polarization gradients and a giant density of non-trivial
ferroelectric domain walls. Additionally, the flexoelectric connection
between polarization and strain gradients is non-trivial in itself, as
it reflects subtle electromechanical couplings, strictly in the long
wavelength limit (${\bf q}\rightarrow {\bf 0}$).\cite{stengel13} All
these are features that our simple MLIPs for {\bto} (trained on DFT
data obtained from small 40-atom cells) cannot be expected to capture
accurately, in principle. At the same time, our models do contain
information about the (short-range) interatomic couplings that are
expected to dominate the energetics of domain walls in these
materials. (See the Discussion section for a comment on long-range
electrostatic couplings.) Hence, one may wonder: is that information
enough to predict, at a qualitative level at least, the occurrence and
main features of the vortex-antivortex pattern of Fig.~\ref{fig:bto}
as obtained from DFT?

We tested our MLIPs by relaxing the vortex-antivortex structures that
some of us studied in Ref.~\onlinecite{sanchezsantolino24} using
DFT. The basic results of this exercise are captured in
Fig.~\ref{fig:bto}, which shows the electric dipoles extracted from
the atomic structure following the same procedure as in
Ref.~\onlinecite{sanchezsantolino24}. (In essence, we plot Ti-centered
electric dipoles, computed for each Ti cation in the lattice.)
Remarkably, both MLIPs relax to local energy minima that retain the
vortex-antivortex texture obtained in DFT and observed
experimentally. The agreement between the structures obtained with DFT
and the VASP MLIP is nearly perfect; the Allegro model, on the other
hand, predicts correctly the magnitude of the local dipoles but fails
to capture the size of the vortex and antivortex cores. A clear
connection between electric texture and local strain is established in
all cases (not shown here). Finally, the energy difference between the
textured state and the competing homogeneous solution with
polarization in-plane is 10~meV/f.u. and 1~meV/f.u, respectively, for
the VASP and Allegro MLIP, both remarkably close to the DFT value of
9~meV/f.u.

In conclusion, our minimal MLIPs for {\bto} perform very well when
applied to a problem that clearly lies beyond the training space
explored in the model construction. There is no doubt that these MLIPs
could have been used to reach the basic conclusions of the DFT
analysis in Ref.~\onlinecite{sanchezsantolino24}, namely, the
existence of low-lying local energy minima featuring a lattice of
electric vortices and antivortices, and the connection between such a
dipole pattern and the spatial modulation of shear strains. The
difference: a 10$^{4}$-fold speed up in the corresponding simulations,
from weeks to minutes of CPU time.

\begin{figure*}[t]
    \centering
    \includegraphics[width=1\linewidth]{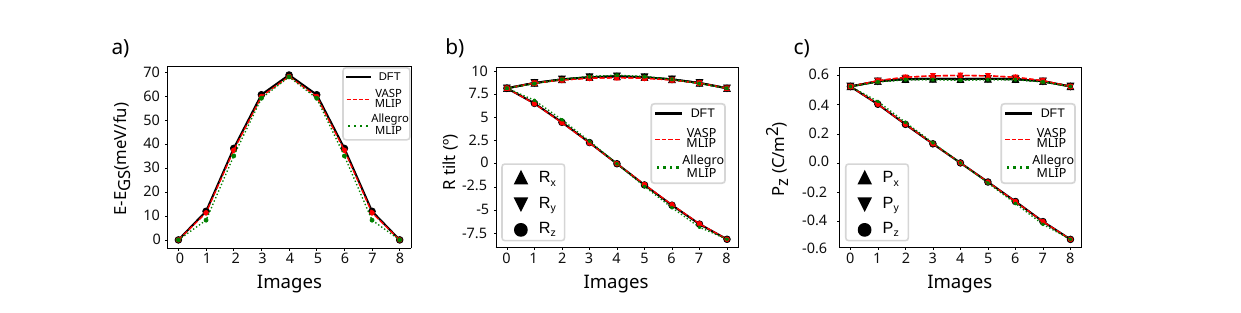}
    \caption{Ferroelectric switching path for {\bfo} (see text). Shown
      are, as a function of structure along the path, the energy vs
      polarization (a), polarization components (b), and antiphase
      tilt components (c).}
    \label{fig:bfo}
\end{figure*}

\subsubsection{BiFeO$_{3}$: Switching path}

So far, {\bfo} is the only room-temperature magnetoelectric
multiferroic material of practical importance.\cite{catalan09} Its
remanent polarization has been experimentally measured to be one of
the largest among ferroelectrics, with $P \approx 1$~C/m$^{2}$, while
the typical coercive field is also relatively large (around 200~kV/cm
for bulky samples). Room-temperature electric control of magnetization
was demonstrated in {\bfo} already in 2014.\cite{heron14} To advance
toward a technological use of such a magnetoelectric switching effect,
several outstanding issues remain, including a needed reduction in the
coercive field so that ultralow-power actuation can be achieved. A
significant research effort focuses on this goal.\cite{prasad20,huang20}

In this context, there is interest in studying switching barriers and
paths in {\bfo}, which can be estimated from DFT, e.g. by using the
so-called ``nudged elastic band'' method.\cite{sheppard12} Critically,
both the switching barrier and the corresponding structural path need
to be optimized in order to achieve ultralow-power magnetoelectric
switching in {\bfo}. Specifically, it is important to work in
conditions such that polarization switching is accompanied by a
reversal of the antiphase octahedral tilts characteristic of {\bfo}'s
ferroelectric phase, as the latter control the sign of the
magnetization.\cite{heron14,ederer05a,rahmedov12}

Hence, to test our {\bfo} models, we compute the switching path
corresponding to the reversal of one polarization component, which
captures the main features of the effect. Figure~\ref{fig:bfo} shows
the comparison between DFT and MLIP results. The MLIPs predict almost
exactly the energy barrier and corresponding distortions, including
the sign reversal of the $R_{z}$ component of the antiphase O$_{6}$
tilts. Bear in mind that the switching proceeds through a transition
state, with $P_{z} = R_{z} = 0$ that is never explored in the training
(see Fig.~\ref{fig:expspac}). Yet, the DFT energy barrier is
reproduced within 1~meV/f.u.

\setlength{\tabcolsep}{8pt}
\begin{table}[t]
  \caption{Comparison of DFT and MLIP energies for the four low-energy
    phases of {\pzo} considered (see text). Energies are given in
    meV/f.u. The AFE$_{40}$ phase is taken as the zero of energy.}
  \vskip 2mm
  \centering
  \begin{tabular}{cccc}
    \hline\hline & DFT & VASP MLIP & Allegro MLIP \\ \hline FiE & -0.84
    & -2.04 & -3.80 \\ AFE$_{80}$ & -0.89 & -4.66 & -1.78
    \\ AFE$_{40}$ & 0.00 & 0.00 & 0.00 \\ FE & +0.23 & -11.12 &
    -13.11 \\ Cubic & 278 & 272 & 261\\ \hline\hline
\end{tabular}
  \label{tab:pzo}
\end{table}

\subsubsection{PbZrO$_{3}$: discovering unprecedented phases}

{\pzo} may well be the inorganic perovskite with the richest
low-energy landscape.\cite{iniguez14,baker21,aramberri21,zhang24}
Experimentally, the bulk stable phase at ambient conditions is
antiferroelectric (AFE$_{40}$ in the following). It features oxygen
octahedral rotations, antiparallel displacements of the Pb cations,
and significantly deformed O$_{6}$ groups. A rhombohedral
ferroelectric phase (FE) with $R3c$ symmetry is very close in energy
and generally assumed to be the structure the material adopts upon
application of an electric field. Then, recent calculations by some of
us predicted that a ferrieletric phase (FiE), with uncompensated
antiparallel displacements of the Pb atoms, has a lower energy than
the AFE$_{40}$ polymorph. Further, it has been shown from first
principles that the AFE$_{40}$ phase itself presents a cell-doubling
instability, yielding yet another low-energy structure
(AFE$_{80}$).\cite{baker21,aramberri21} Strikingly, DFT predicts that
these four phases lie within a 1~meV/f.u. energy
window~\cite{aramberri21}, which poses a challenge for effective
modeling.

Table~\ref{tab:pzo} shows our results for the energies of {\pzo}'s
main competing polymorphs. (The MLIP-obtained lattice parameters and
atomic structures are virtually identical to the DFT results, with
deviations of 1~\% at most, and are not shown here.) Most remarkably,
our MLIPs correctly predict the existence of the AFE$_{40}$,
AFE$_{80}$, and FiE phases. Further, these polymorphs are found to lie
within a few meV per formula unit of each other, essentially capturing
the near-degeneracy obtained from DFT. Table~\ref{tab:pzo} also shows
that our minimal MLIPs are not perfect, though, as they exaggerate the
stability of the FE polymorph. Further, the energy gap between the
cubic perovskite structure and the low-energy polymorphs is severely
underestimated.

To see this result in its appropriate context, note that our MLIPs for
{\pzo} are constructed based on knowledge about common and simple
perovskites, i.e., phases displaying standard polar distortions and
``Glazer patterns'' of perfect in-phase and antiphase O$_{6}$
octahedral tilts.\cite{glazer75} Our exercise shows that such MLIPs
would have allowed us to discover exceedingly rare and non-trivial new
phases, such as the mentioned AFE$_{40}$, AFE$_{80}$, and FiE
structures, that are nowhere to be found in the data used to construct
the models. More specifically, our MD simulations explore distorted
low-energy configurations that lie far -- both structure and energy
wise -- from the perfect cubic structure. More critically, our minimal
training set is derived from DFT simulations using a 40-atom supercell
that can be seen as a $2\times 2\times 2$ repetition of the elemental
5-atom perovskite unit, which is incompatible with any of the AFE and
FiE polymorphs that attract so much attention today. Nevertheless:
Based on the MLIP predictions, we could then refine the models through
further training on the new polymorphs, so that, for example, the
energy gap with the FE structure in Table~\ref{tab:pzo} be corrected.

Our {\pzo} example thus suggests that MLIPs provide us with a platform
for materials discovery studies, starting with minimal models that can
eventually be improved based on the most surprising predictions coming
from the models themselves.

\setlength{\tabcolsep}{8pt}
\begin{table}[t]
  \caption{Comparison of DFT and MLIP energies for the key polymorphs
    of {\hfo} considered (see text). Energies are given in
    meV/f.u. The monoclinic phase is taken as the zero of energy.}
  \vskip 2mm
  \centering
  \begin{tabular}{cccc}
    \hline\hline
    & PBEsol & VASP MLIP &  Allegro MLIP  \\ \hline
  monoclinic &    0    &  0   &   0 \\
  tetragonal &  139    & 140  & 143 \\
  oIII       &   64    & 65   &  67 \\
  oI         &   16    &  14  &  60 \\ 
  cubic      &  207    & 188  & 208 \\ \hline\hline
\end{tabular}
\label{tab:hfo}
\end{table}

\subsubsection{HfO$_{2}$: polymorphism and ferroelectricity}

Of the ferroelectric oxides in this study, {\hfo} can be considered
the odd one out. Prior to the discovery of its polar phase in
2011,\cite{boscke11} {\hfo} was already industrially relevant, in use
as a high-k gate dielectric in field-effect transistors. Consequently,
the observation of ferroelectricity in ultrathin films immediately
drew enormous interest. Several MLIP studies of {\hfo} and the closely
related fluorite ZrO$_{2}$ have been reported, typically based on deep
neural network potentials.\cite{wu21} Applications of such MLIPs
include studies of oxygen mobility,\cite{ma23} piezo- and pyroelectric
responses,\cite{ganser22} or the mechanism of antiferroelectricity in
ZrO$_{2}$.\cite{ganser24}

Within our minimalist approach, we aim at (re)discovering the most
common ferroelectric phase of {\hfo} (usually denoted ``oIII'', with
orthorhombic symmetry $Pca2_1$). We also examine whether our models
predict the correct hierarchy between competing polar and non-polar
polymorphs. Our results are summarized in Table~\ref{tab:hfo}. (We
only provide energies in this Table, as the MLIP-predicted structures
match the DFT results to within 1\% for all considered phases.)
Remarkably, our MLIPs yield a stable oIII polymorph with nearly
perfect energy as compared to the DFT result. The excellent accuracy
in the oIII prediction can be attributed to the fact that our MD runs
find this state serendipitously (see Section~\ref{sec:mlip} and
Fig.~\ref{fig:expspac}), so it is in fact included in the training
data. More surprising is the fact that both MLIPs reproduce correctly
the ordering, energy wise, of all the polymorphs considered. While we
find some significant quantitative deviations -- as e.g. in the
VASP-MLIP result for the cubic phase, or the Allegro-MLIP result for
the oI polymorph -- those were to be expected, as the cubic and oI
states are nowhere to be found in the training space.

It is also worth noting that an even simpler MLIP (trained on DFT data
for 12-atom cells, and comprising only 63 DFT calculations) is already
capable of predicting the oIII local minimum with near-DFT accuracy in
what regards the structure and a remarkably solid value for the energy
(42~meV/f.u. above the monoclinic ground state vs. 64~meV/f.u. from
DFT). Another unique feature of oIII hafnia is that it features a
negative energy of its 180$^{\circ}$ ferroelectric domain walls. This
striking result stems from the fact that the antipolar oI state (which
can be viewed as a ferroelectric multidomain configuration) lies below
the oIII structure, a property predicted by our MLIPs.

\section{Discussion}\label{sec:discussion} 

Our results carry some implications -- both general and specific to
ferroelectrics -- that merit a brief comment.

This study provides evidence for how relatively simple models can
reliably predict phenomena outside their theoretical application
scope. Hence, our results can be viewed in the context of a broader
theme: how increasing model complexity can degrade predictive power
outside the training range.\cite{nakkiran19,gulrajani20} In the
future, it would be interesting -- and potentially rewarding -- to
explore how recent ``domain generalization'' approaches perform in
materials-science problems, e.g. those involving the family of
ferroelectric materials considered here. Conversely, it may be
valuable to examine whether domain expertise in ferroelectrics and
related condensed-matter fields could help define testbeds where the
predictive power of general ML schemes can be evaluated, thereby
contributing to the development of smarter and more efficient
general-purpose algorithms.

Along the same lines, our results suggest that it would be worthwhile
to conduct a more systematic investigation into how, and under what
circumstances, excessive training of ML models -- such as general
neural networks -- may lead to reduced predictive power outside the
training space. Likewise, our findings motivate further studies on the
predictive performance of MLIPs even simpler than those considered
here; for example, families of interpretable potentials based on
physically motivated functional forms.\cite{fan21,xie23} In our view,
striking the right balance between accuracy and predictive power will
be essential for establishing practical usage guidelines for MLIPs and
developing protocols tailored to the goals of specific
investigations. MLIPs offer almost unlimited flexibility in model
choice for particular applications, and making informed decisions on
how to proceed in each case may increasingly become the distinctive
human contribution to this field. Achieving that will require
systematic studies that build upon the somewhat anecdotal -- though,
in our view, very compelling -- evidence provided by works such as the
present one.

When it comes to the particular case of the ferroelectrics
investigated in this work, we should first stress that the incredible
predictions reported here are a testimony to the value of the employed
MLIP and on-the-fly learning
schemes,\cite{bartok15,jinnouchi19,bartok13,musaelian23,tan25,batzner22}
which enable the construction of incredibly useful models based on a
very modest amount of DFT simulations. This is remarkable particularly
when it comes to neural network potentials, commonly believed to be
exceedingly data hungry.

It is also worth noting that this project began as a test of a very
specific hypothesis. We believed that kernel-based GAP-SOAP MLIPs
should be able to capture the dominant short-range interactions in
ferroelectric oxides from a modest amount of training data. In
particular, we aimed to explore the feasibility of using very small
simulation supercells for training. Our intuition was based on earlier
work with second-principles potentials,\cite{wojdel13,zubko16} which
showed that DFT data from small supercells is sufficient to
characterize the interatomic interactions underlying remarkably
complex structural phenomena -- such as the electric skyrmions
predicted (and later experimentally confirmed) in perovskite
ferroelectrics.\cite{goncalves19,das19,wojdel14a} The present study
confirms that initial hypothesis and, more unexpectedly, indicates
that our intuition extends to curated neural-network frameworks such
as Allegro. Ultimately, we think this serves as an example of how a
physics-based expectation -- namely, that relatively simple
short-ranged couplings can produce great complexity, yet remain easy
to learn and model -- can guide the identification of promising
strategies for using MLIPs.

As a final note, long-range Coulomb interactions between electric
dipoles are known to play a major role in the physics of
ferroelectrics, as e.g. they create depolarizing fields that are one
of the main factors leading to the formation of complex multidomain
structures. Traditionally, effective atomistic potentials of
ferroelectrics have treated such dipole-dipole interactions exactly,
using their well-known analytic form and evaluating them via Ewald
sums.\cite{gonze97,zhong95a,wojdel13} The difficulties to include such
interactions within MLIP schemes was a blocking point for the
widespread adoption of the new models in the ferroelectrics
community.\cite{zhang22} Nevertheless, growing evidence suggest that
MLIPs including mid-ranged interactions (e.g., within 5~\AA\ to 8~\AA,
as done here) are able to effectively capture all relevant couplings,
and that a careful treatment of the long-range Coulomb coupling is
only needed to address very specific features (e.g., for a correct
quantitative treatment of non-analytical properties such as the
longitudinal-transversal optical frequency
splitting).\cite{yang24,monacelli24,yu25} The present results further
support this conclusion and the usefulness of short-ranged MLIPs to
study complex ferroelectric phenomena.

\section{Summary and conclusions}\label{sec:summary}

In this work we examine the performance of machine-learned interatomic
potentials (MLIPs) in predicting intricate structural (and, to some
extent, functional) properties of representative ferroelectric and
related materials. We work with standard and widely-available
kernel-based models, as well as with neural network models derived
from the same training sets. We restrict ourselves to small training
sets generated on-the-fly and without any fine tuning. We also adopt
modest unoptimized choices of the MLIP-defining hyperparameters. Then,
we test the predictive performance of the resulting minimalist MLIPs.

Within the explored parameter space, both types of MLIPs reproduce DFT
energies, forces, and stresses with good fidelity. More notably, the
models extrapolate in a physically meaningful way beyond the training
space, correctly predicting critical, previously unexpected properties
of these compounds -- from the emergence of novel polymorphs and
topological states to the occurrence of complex polarization switching
paths. The success of these minimalist MLIPs challenges the prevailing
view that reliable potentials necessarily require large, curated
datasets and extensive optimization. Rather, our results show that
simple, low-cost and low-knowledge training strategies can yield
models capable of unveiling unprecedented non-trivial behaviors. If
this is true for materials as complex -- structurally and
lattice-dynamically -- as the considered ferroelectric oxides, it is
essentially guaranteed that our conclusions will pertain to many other
materials families as well.

Our results also suggest that complex structural problems such as
those considered here could serve as valuable testbeds for
domain-generalization approaches currently being explored in the
context of general machine learning. Our work also opens promising
avenues for investigating the use of simple MLIPs in exploratory
research explicitly aimed at uncovering novel ``out-of-scope''
phenomena. We thus hope that the present work will contribute to a
more informed and effective use of ML methods, broadening their role
from tools of interpolation to practical instruments for discovering
and predicting emergent physical behavior.

Fruitful discussions with N. Bristowe (Durham) are gratefully
acknowledged. Work supported by the Luxembourg National Research Fundt
through grants BRIDGES/18421428/SWITCHON (I.R.-M. and J.\'I.-G.,
co-funded by Intel Corporation), INTER/NSF/24/18804122/PIEZOHAFNIA
(B.M. and J.\'I.-G.), and C23/MS/17909853/BUBBLACED (J.\'I.-G. and
H.A.).

\bibliography{biblio}

\end{document}